# Branching ratios of $B_c$ Meson Decays into Tensor Meson in the Final State

Neelesh Sharma
Department of Physics, Punjabi University,
Patiala-147 002, INDIA.
E-mail: nishu.vats@gmail.com

**Abstract**

Two-body hadronic weak decays of $B_c$ meson involving tensor meson in the final state are studied by using Isgur-Scora-Grinstein-Wise (ISGW II) model. Decay amplitudes are obtained using the factorization scheme in the Spectator Quark Model. Branching ratios for the charm changing and bottom changing decay modes are predicted.

## I. INTRODUCTION

Studies of the $B_c$ meson decays are important for several reasons. The $B_c$ meson discovered at Fermilab [1] is the only quark-antiquark bound system $(\bar{b}c)$ composed of heavy quarks $(b,c)$ with different flavors, and are thus flavor asymmetric. Recently, CDF Collaboration [2] announced an accurate determination of the $B_c$ meson mass and its life time. The investigation of the $B_c$ meson properties (mass spectrum, decay rates, etc.) is therefore of special interest compared to symmetric heavy quarkonium $(\bar{b}b,\bar{c}c)$ states. Also, $B_c$ meson attracts the interest of experimentalists for testing the predictions of perturbative QCD in the laboratory. The difference of quark flavors forbids the annihilation of $B_c$ meson into gluons. As a result, the pseudoscalar $(\bar{b}c)$ state is much more stable than the heavy quarkonium states, and decays only weakly. The decay processes of the $B_c$ meson can be broadly divided into two classes: involving the decay of $b$ quark, and that of $c$ quark, besides the annihilation of $b$ and $\bar{c}$. Preliminary estimates of the widths of some decay channels of $B_c$ have been made to show that the bound state effects may be significant in $B_c$ decays. Experimental study of the $B_c$ mesons are in plan for B-Physics both at the TEVATRON and Large Hadron Collider (LHC).

Earlier, there is a lot of work done in the semileptonic and nonleptonic [3-12] decays of the meson $B_c$ to a $s$-wave mesons. Also, the $p$-wave emitting decays of $B_c$ meson have been considered previously by other authors [13,14]. The present work consists the analysis of two-body hadronic weak decays of $B_c$ meson to pseudoscalar ($P$)/vector ($V$) and tensor ($T$) mesons in the final state employing the factorization hypothesis. The CKM-favored modes in charm changing and bottom changing decay modes are calculated using the non-relativistic quark model proposed by Isgur-Scora-Grinstein-Wise (ISGW) [15].

The present paper is organized as follows. Sec. II, includes meson spectroscopy. In Sec. III, methodology for $B_c \to PT/VT$ is discussed and the form factors involving $B_c \to T$ transition are calculated. Sec. IV deals with numerical results and discussions and summary and conclusions are given in the last section.

## II. MESON SPECTROSCOPY

Experimentally [16], the tensor meson sixteen-plet comprises of an isovector $a_2(1.318)$, strange isospinor $K_2^*(1.429)$, charm SU(3) triplet $D_2^*(2.457)$, $D_{s2}^*(2.573)$ and three isoscalars $f_2(1.275)$, $f_2'(1.525)$ and $\chi_{c2}(3.555)$. These states behave well with respect to the quark model assignments, though the spin and parity of the charm isosinglet $D_{s2}^*(2.573)$ remain to be confirmed. The numbers given within parentheses indicates the mass (in GeV units) of the respective mesons. $\chi_{c2}(3.555)$ is assumed to be pure $(c\bar{c})$ state, and mixing of the isoscalar states is defined as:

$$f_2(1.275) = \frac{1}{\sqrt{2}}(u\bar{u} + d\bar{d})\cos\phi_T + (s\bar{s})\sin\phi_T,$$

$$f_2^{'}(1.525)\frac{1}{\sqrt{2}}(u\bar{u} + d\bar{d})\sin\phi_T - (s\bar{s})\cos\phi_T, \tag{1}$$

where $\phi_T = \theta(ideal) - \theta_T(physical)$ and $\theta_T(physical) =$ $27^\circ$.

For $\eta$ and $\eta'$ states of well established pseudoscalar sixteen-plet, we use

$$\eta(0.547) = \frac{1}{\sqrt{2}}(u\bar{u} + d\bar{d})\sin\phi_P - (s\bar{s})\cos\phi_P,$$

$$\eta'(0.958) = \frac{1}{\sqrt{2}}(u\bar{u} + d\bar{d})\cos\phi_P + (s\bar{s})\sin\phi_P, \tag{2}$$

where $\phi_p = \theta(ideal) - \theta_p(physical)$ and we take $\theta_P(physical) = -15.4^\circ$ [16]. $\eta_c$ is taken as

$$\eta_c(2.979) = (c\bar{c}). \tag{3}$$

Similarly, for $\omega$ and $\phi$ states of well established pseudoscalar sixteen-plet, we use

$$\omega(0.783) = \frac{1}{\sqrt{2}}(u\bar{u} + d\bar{d})\cos\phi_V + (s\bar{s})\sin\phi_V,$$

$$\phi(1.019) = \frac{1}{\sqrt{2}}(u\bar{u} + d\bar{d})\sin\phi_V - (s\bar{s})\cos\phi_V, \tag{4}$$

where $\phi_V = \theta(ideal) - \theta_V(physical)$ and we take $\theta_V(physical) = 39^\circ$ [16]. $J/\psi$ is taken as

$$J/\psi(3.097) = (c\bar{c}). \tag{5}$$

## III. METHODOLOGY

### A. WEAK HAMILTONIAN

To the lowest order in weak interaction, the non-leptonic Hamiltonian has the usual current $\otimes$ current form

$$H_w = \frac{G_F}{\sqrt{2}} J^{+}{}_{\mu} J^{\mu} + h.c. \tag{6}$$

The weak current $J_\mu$ is given by

$$J_\mu = (\bar{u}\ \bar{c}\ \bar{t}\ )\,\gamma_\mu(1\text{-}\gamma_5)\begin{pmatrix} d' \\ s' \\ b' \end{pmatrix}, \tag{7}$$

where $d'$, $s'$ and $b'$ are mixture of the $d, s$ and $b$ quarks, as given by Cabibbo-Kobayashi-Maskawa (CKM) matrix [16].

### a) For bottom changing decays

The QCD modified weak Hamiltonian [17] generating the $b$ quark decays in CKM enhanced modes ($\Delta b = 1$, $\Delta C = 1$, $\Delta S = 0$; $\Delta b = 1$, $\Delta C = 0$, $\Delta S = -1$) is given by

$$H_w^{\Delta b=1} = \frac{G_F}{\sqrt{2}}\{V_{cb}V_{ud}^*[c_1(\mu)(\bar{c}b)(\bar{d}u) + c_2(\mu)(\bar{c}u)(\bar{d}b)] + V_{cb}V_{cs}^*[c_1(\mu)(\bar{c}b)(\bar{s}c) + c_2(\mu)(\bar{c}c)(\bar{s}b)]\}, \tag{8}$$

where $G_F$ is the Fermi constant and $V_{ij}$ are the CKM matrix elements, $c_1$ and $c_2$ are the standard perturbative QCD coefficients.

### b) For bottom conserving and charm changing decays

In addition to the bottom changing decays, bottom conserving decay channel is also available for the $B_c$ meson, where the charm quark decays to an $s$ or $d$ quark. The weak Hamiltonian generating the $c$ quark decays in CKM enhanced mode ($\Delta b = 0$, $\Delta C = -1$, $\Delta S = -1$) is given by

$$H_w^{\Delta c=-1} = \frac{G_F}{\sqrt{2}} V_{ud} V_{cs}^* [c_1(\mu)(\bar{u}d)(\bar{s}c) + c_2(\mu)(\bar{u}c)(\bar{s}d)]. \tag{9}$$

One naively expects this channel to be suppressed kinematically due to the small phase space available. However, the kinematic suppression is well compensated by the CKM element $V_{cs}$, which is larger than $V_{cb}$ appearing for the bottom changing decays. In fact, we shall show later that bottom conserving decay modes are more prominent than the bottom changing ones.

By factorizing matrix elements of the four-quark operator contained in the effective Hamiltonian (8) and (9), one can distinguish three classes of decays [18]:

- The first class contains those decays which can be generated from color singlet current and the decay amplitudes are proportional to $a_1$, where $a_1(\mu) = c_1(\mu) + \frac{1}{N_c} c_2(\mu)$, and $N_c$ is the number of colors.
- Second classes of transitions consist of those decays which can be generated from neutral current. The decay amplitude in this class is proportional to $a_2$ i.e. for the color suppressed modes $a_2(\mu) = c_2(\mu) + \frac{1}{N_c} c_1(\mu)$.
- The third class of decay modes can be generated from the interference of color singlet and color neutral currents i.e. the $a_1$ and $a_2$ amplitudes interfere.

Following the convention of large $N_c$ limit to fix QCD coefficients $a_1 \approx c_1$ and $a_2 \approx c_2$, where [18] $c_1$ and $c_2$ are fixed as:

$$
\begin{aligned}
&c_1(\mu) = 1.26 \ , \ c_2(\mu) = -0.51 \ \text{at} \ \mu \approx m_c^2 , \\
&c_1(\mu) = 1.12 \ , \ c_2(\mu) = -0.26 \ \text{at} \ \mu \approx m_b^2 .
\end{aligned} \tag{10}
$$

## B. DECAY AMPLITUDES AND RATES

### a) $B_c \to PT$ Decay:

The decay rate formula for $B_c \to PT$ decays is given by

$$
\Gamma(B_c \to PT) = \left( \frac{m_{B_c}}{m_T} \right)^2 \frac{p_c^5}{12\pi m_T^2} \left| A(B_c \to PT) \right|^2 , \tag{11}
$$

where $p_c$ is the magnitude of the three-momentum of a final-state particle in the rest frame of $B_c$ meson and $m_T$ denotes the mass of the tensor meson.

The factorization scheme expresses the decay amplitude as the product of matrix elements of weak currents (up to the weak scale factor of $\frac{G_F}{\sqrt{2}} \times$ CKM elements×QCD factor) as

$$
\langle PT | H_W | B_c \rangle \sim \langle P | J^{\mu} | 0 \rangle \langle T | J_{\mu} | B_c \rangle + \langle T | J^{\mu} | 0 \rangle \langle P | J_{\mu} | B_c \rangle , \tag{12}
$$

However, the matrix element

$$\langle T | J^{\mu} | 0 \rangle = 0 , \tag{13}$$

because the trace of the polarization tensor $\in^{\mu\nu}$ of the tensor meson $T$ vanishes and the auxiliary condition holds, $p_T^{\mu} \in^{\mu\nu} = 0$. Thus, in the generalized factorization scheme, the decay amplitudes for $B_c \to PT$ are considerably simple in comparison to the other two-body charmless decays of $B$ mesons.

One can use the ISGW quark model [15] to analyze two-body decay processes $B_c \to PT$ in the framework of generalized factorization, where the parameterizations of the hadronic matrix elements in $B_c \to PT$ decays are described as:

$$\langle P | J^{\mu} | 0 \rangle = -if_P k_{\mu} ,$$

$$\begin{aligned} \langle T | J^{\mu} | B_c \rangle = {} & ih \in_{\mu\upsilon\lambda\rho} \in^{*\upsilon\alpha} P_{B_c\alpha} (P_{B_c} + P_T)^{\lambda} (P_{B_c} - P_T)^{\rho} \\ & + k \in^{*}_{\mu\upsilon} P^{\upsilon}_{B_c} + b_+ (\in^{*}_{\alpha\beta} P^{\alpha}_{B_c} P^{\beta}_{B_c}) [(P_{B_c} + P_T)_{\mu} + b_- (P_{B_c} - P_T)_{\mu} ], \end{aligned} \tag{14}$$

where $k_{\mu} = (P_B - P_T)_{\mu}$ and $f_P$ denotes the decay constant of pseudoscalar mesons. $P_{B_c}$ and $P_T$ denotes the momentum of the $B_c$ meson and the tensor meson, respectively. The ISGW model yields

$$\langle PT | H_W | B_c \rangle = -if_P (\in^{*}_{\mu\upsilon} P^{\mu}_{B_c} P^{\upsilon}_{B_c}) F^{B_c \to T} (m_P^2), \tag{15}$$

where

$$F^{B_c \to T} = k + (m_{B_c}^2 - m_T^2) b_+ + m_P^2 b_- . \tag{16}$$

$\in^{*}_{\mu\nu}$ is the symmetric and traceless tensor describing the polarization of tensor mesons. The argument, $F^{B_c \to T}$, in the function means that the form factors $k$, $b_+$ and $b_-$ should be evaluated at $m_P^2$.

Thus the decay amplitude, in general, have the following form

$$A(B \to PT) = \frac{G_F}{\sqrt{2}} \times (CKM\ factors \times QCD\ factors) \times f_P F^{B_c \to T} (m_P^2) . \tag{17}$$

**b) $B_c \to VT$ Decay:**

The decay rate formula for $B_c \to VT$ [19] is,

$$\Gamma(B_c \to VT) = \frac{G_F^2}{48\pi m_T^4} m_V f_V^2 [\alpha|\vec{p}_V|^7 + \beta|\vec{p}_V|^5 + \gamma|\vec{p}_V|^3], \qquad (18)$$

where $|p_V|$ is the magnitude of the three-momentum of the final-state particle $V$ or $T$ ($|p_V| = |p_T|$) in the rest frame of $B_c$ meson. $\alpha$, $\beta$ and $\gamma$, respectively, are quadratic functions of the form factors, are given by

$$\alpha = 8m_{B_c}^4 b_+^2,$$
$$\beta = 2m_{B_c}^2 [6m_V^2 m_T^2 h^2 + 2(m_{B_c}^2 - m_T^2 - m_V^2)kb_+ + k^2], \qquad (19)$$
$$\gamma = 5m_T^2 m_V^2 k^2.$$

Here also the decay amplitude can be expressed as the product of matrix elements of weak currents (up to the weak scale factor of $\frac{G_F}{\sqrt{2}} \times$ CKM elements×QCD factor):

$$\langle VT|H_W|B_c\rangle \sim \langle V|J^\mu|0\rangle\langle T|J_\mu|B_c\rangle + \langle T|J^\mu|0\rangle\langle V|J_\mu|B_c\rangle, \qquad (20)$$

The hadronic matrix elements is:

$$\langle V|J_\mu|0\rangle = \in_\mu^* m_V f_V, \qquad (21)$$

where $f_V$ denotes the decay constant of the vector meson. Relations (14) and (21) yields

$$\langle VT|H_W|B_c\rangle = m_V f_V F^{B_c \to T}(m_V^2), \qquad (22)$$

where

$$F_{\alpha\beta}^{B_c \to T} = \in_\mu^* (P_{B_c} + P_T)_\rho [ih\varepsilon^{\mu\nu\rho\sigma} g_{\alpha\nu}(P_V)_\beta (P_V)_\sigma + k\delta_\alpha^\mu \delta_\beta^\rho + b_+(P_V)_\alpha (P_V)_\beta g^{\mu\rho}]. \qquad (23)$$

In the above expressions, $\in_\mu^*$ denotes the polarization four-vector of $V$. The form factors $h$, $k$, $b_+$ and $b_-$ should be evaluated at $m_V^2$.

Finally,

$$A(B \to VT) = \frac{G_F}{\sqrt{2}} \times (CKM\ factors \times QCD\ factors) \times m_V f_V \in^{*\alpha\beta} F_{\alpha\beta}^{B_c \to T}(m_V^2). \qquad (24)$$

The required form factors $h$, $k$, $b_+$ and $b_-$ are calculated from the following expressions of ISGW II model [15]:

$$h = \frac{m_d}{2\sqrt{2\tilde{m}_{B_c}\beta_{B_c}}}\left(\frac{1}{m_q} - \frac{m_d\beta_{B_c}^2}{2\mu_-\tilde{m}_T\beta_{B_cT}^2}\right)F_5^{(h)},$$

$$k = \frac{m_d}{\sqrt{2\beta_{B_c}}}(1+\tilde{\omega})\; F_5^{(k)},$$

$$b_+ + b_- = \frac{m_d^2}{4\sqrt{2m_q m_b \tilde{m}_{B_c}\beta_{B_c}}}\frac{\beta_T^2}{\beta_{B_cT}^2}\left(1 - \frac{m_d}{2\tilde{m}_{B_c}}\frac{\beta_T^2}{\beta_{B_cT}^2}\right)F_5^{(b_+ + b_-)}, \tag{25}$$

$$b_+ - b_- = -\frac{m_d}{\sqrt{2m_b\tilde{m}_T\beta_{B_c}}}\left(1 - \frac{m_d m_b}{2\mu_+\tilde{m}_{B_c}}\frac{\beta_T^2}{\beta_{B_cT}^2} + \frac{m_d}{4m_q}\frac{\beta_T^2}{\beta_{B_cT}^2}\left(1 - \frac{m_d}{2\tilde{m}_{B_c}}\frac{\beta_T^2}{\beta_{B_cT}^2}\right)\right)F_5^{(b_+ - b_-)},$$

where

$$F_5^{(h)} = F_5\left(\frac{\bar{m}_{B_c}}{\tilde{m}_{B_c}}\right)^{-3/2}\left(\frac{\bar{m}_T}{\tilde{m}_T}\right)^{-1/2},$$

$$F_5^{(k)} = F_5\left(\frac{\bar{m}_{B_c}}{\tilde{m}_{B_c}}\right)^{-1/2}\left(\frac{\bar{m}_T}{\tilde{m}_T}\right)^{1/2}, \tag{26}$$

$$F_5^{(b_+ + b_-)} = F_5\left(\frac{\bar{m}_{B_c}}{\tilde{m}_{B_c}}\right)^{-5/2}\left(\frac{\bar{m}_T}{\tilde{m}_T}\right)^{1/2},$$

$$F_5^{(b_+ - b_-)} = F_5\left(\frac{\bar{m}_{B_c}}{\tilde{m}_{B_c}}\right)^{-3/2}\left(\frac{\bar{m}_T}{\tilde{m}_T}\right)^{-1/2},$$

and

$$\tilde{\omega} - 1 = \frac{t_m - t}{2\bar{m}_{B_c}\bar{m}_T}. \tag{27}$$

The common scale factor is given by

$$F_n = \left(\frac{\tilde{m}_T}{\tilde{m}_{B_c}}\right)^{1/2}\left(\frac{\beta_T\beta_{B_c}}{\beta_{B_cT}^2}\right)^{5/2}\left[1 + \frac{1}{18}\chi^2(t_m - t)\right]^{-3}, \tag{28}$$

where

$$\chi^2 = \frac{3}{4m_b m_q} + \frac{3m_d^2}{2\bar{m}_{B_c}\bar{m}_T\beta_{B_cT}^2} + \frac{1}{\bar{m}_{B_c}\bar{m}_T}\left(\frac{16}{33-2n_f}\right)\ln\left[\frac{\alpha_S(\mu_{QM})}{\alpha_S(m_q)}\right], \tag{29}$$

and

$$\beta_{B_cT}^2 = \frac{1}{2}\left(\beta_{B_c}^2 + \beta_T^2\right), \tag{30}$$

$\tilde{m}$ is the sum of the mesons constituent quark masses, $\bar{m}$ is the hyperfine averaged physical masses, $n_f$ is the number of active flavors, which is taken to be five in the present case, $t_m = (m_{B_c} - m_T)^2$ is the maximum momentum transfer and

$$\mu_+ = \left(\frac{1}{m_q} + \frac{1}{m_b}\right)^{-1}, \tag{31}$$

with $m_q$ and $m_d$ being the masses of the quark $q_1$ and $\bar{q}_2$, respectively. Following ISGW II quark model values [15] of the quark masses (in GeV) are used:

$$m_u = m_d = 0.33,\ m_s = 0.55,\ m_c = 1.82,\ m_b = 5.20, \tag{32}$$

and values of the parameter $\beta$'s for different $s$-wave and $p$-wave mesons as given in the Table I. The obtained the form factors describing $B_c \to T$ transitions are given in Table II at $q^2 = t_m$. It may be noted that $B_c \to P/V$ form factors do not appear in the decay amplitudes given in Tables III and IV. However, in order to get an estimate of these form factors and the predictability of the model, one can compare the results of semileptonic weak decays for $B_c \to P/V$ channels in ISGW II model [15] with the prediction of other models as shown in Table V, we find that the results are reasonably comparable with the other models.

## IV. NUMERICAL RESULTS AND DISCUSSIONS

Sandwiching the weak Hamiltonian (8) and (9) between the initial and final states, we obtain the decay amplitudes of $B_c$ meson for the various decay modes as given in the Tables III and IV. For numerical calculations, we use the following values of the decay constants (given in GeV units) of pseudoscalar mesons [8, 16] and vector mesons [17]:

$$f_\pi = 0.131\ ,\ f_K = 0.160\ ,\ f_D = 0.223\ ,\ f_{D_s} = 0.294\ ,$$
$$f_\eta = 0.133\ ,\ f_{\eta'} = 0.126\ ,\ f_{\eta_c} = 0.400\,. \tag{33}$$

and

$$f_\rho = 0.221\ ,\ f_{K^*} = 0.220\ ,\ f_{D^*} = 0.245,\ f_{D_s^*} = 0.273\ ,$$

$$f_\omega = 0.195,\ f_\phi = 0.229\ ,\ f_{J/\psi} = 0.411. \tag{34}$$

Finally, branching ratios of $B_c \to PT$ meson decays in charm changing and bottom changing decay modes and branching ratios of $B_c \to VT$ meson decays in bottom changing decay modes are calculated. The measurement of these decays would provide an additional test of the quark models used to compute the hadronic matrix elements. The results are given in Tables VI and VII for the various possible Cabibbo-favored decay modes and the observation are listed as follows:

I) For $B_c \to PT$ meson decays (in Table VI):

i) Dominant decays for bottom changing decay modes are, $B(B_c^- \to D_s^- \chi_{c2})$ = 3.2 × 10$^{-4}$ and $B(B_c^- \to \pi^- \chi_{c2})$ = 2.0×10$^{-4}$, which seems to be at the reach of future experiments. The next order dominant decays are $B(B_c^- \to K^- \chi_{c2})$ = 1.5×10$^{-5}$ , $B(B_c^- \to \eta_c D_{s2}^-)$ = 1.4×10$^{-5}$ and $B(B_c^- \to D^- \chi_{c2})$ = 1.2×10$^{-5}$.

ii) Branching ratio of decay, $B(B_c^- \to \pi^- \chi_{c2})$ = 2.0×10$^{-4}$ , are comparable with the numerical value of the recent work [13].

iii) Branching ratio of dominant decay for charm changing decay mode is, $B(B_c^+ \to \pi^+ B_{s2}^0)$ = 3.0×10$^{-4}$, which proceeds via *b* quark as an spectator, has a similar order of branching ratio than *c* quark spectator decays i.e. $B_c^- \to D_s^- \chi_{c2} / \pi^- \chi_{c2}$, although it is suppressed by phase space but favored by the CKM factor.

iv) Among $\Delta b = 1, \Delta C = 1, \Delta S = 0$ mode, $B_c^- \to K^0 K_2^- / \pi^0 a_2^- / \pi^- a_2^0 / \pi^- f_2$ $/ \pi^- f_2' / \eta a_2^- / K^- K_2^0 / \eta' a_2^- / D^- D_2^0 / \eta_c a_2^-$ are forbidden in our analysis. However, these decays occur through the annihilation mechanism. Decay $B_c^- \to D^- D_2^0$ may also be generated through elastic final state interactions (FSIs).

v) In case of $\Delta b = 1, \Delta C = 0, \Delta S = -1$ decay mode, $B_c^- \to \bar{K}^0 D_2^- / D^- \bar{K}_2^0$ $/ \bar{D}^0 K_2^- / D_s^- a_2^0 / D_s^- f_2 / D_s^- f_2'$ are forbidden. However, these decays occur through the annihilation mechanism. Decay $B_c^- \to \bar{K}^0 D_2^-$ may also be generated through elastic final state interactions (FSIs).

II) For $B_c \to VT$ meson decays (in Table VII):

i) For Bottom changing decay modes, branching ratios of dominant mode are $B(B_c^- \to D_s^{*-} \chi_{c2})$ = 8.8×10$^{-4}$ and $B(B_c^- \to \rho^- \chi_{c2})$ = 2.4× 10$^{-4}$. Branching ratio of the decay $B_c^- \to \rho^- \chi_{c2}$ matches well with the value of Ref [14]. Branching ratio for next order dominant decay is $B(B_c^- \to J/\psi D_{s2}^-) = 9.6\times10^{-5}$. Note that the $b$ spectator decays do not appear as they are kinematically forbidden.

ii) In $\Delta b = 1, \Delta C = 1, \Delta S = 0$ mode, $B_c^- \to K^{*0} K_2^- / \rho^0 a_2^- / \rho^- a_2^0 / \rho^- f_2 / \rho^- f_2' / \omega a_2^- / K^{*-} K_2^0 / \phi a_2^- / D^{*-} D_2^0 / J/\psi a_2^-$ are forbidden in the present analysis. However, like $B_c \to PT$ decays these may also occur through the annihilation mechanism. Decay $B_c^- \to D^{*-} D_2^0$ may also be generated through elastic final state interactions (FSIs).

iii) Also, in $\Delta b = 1, \Delta C = 0, \Delta S = -1$ mode, $B_c^- \to \bar{K}^{*0} D_2^- / D^{*-} \bar{K}_2^0 / D_s^{*-} a_2^0 / \bar{D}^{*0} K_2^- / D_s^{*-} f_2 / D_s^{*-} f_2'$ are forbidden and occur through the annihilation mechanism. Decay $B_c^- \to \bar{K}^{*0} D_2^-$ may also be generated through elastic final state interactions (FSIs).

For the sake of comparison, the results of other works [13, 14] are given in the Tables VI and VII. C.H. Chang *et al.* [13] have calculated only the $c$ spectator decay modes using generalized instantaneous approximation, in which branching ratio of $B_c^- \to D_s^{*-} \chi_{c2}$ decay is large as compare to the present value. In general, the present branching ratios of few decays are of the same order of magnitude as observed in [13,14] and in other cases branching ratios are larger as compared to [14]. Ivanov *et al.* [8] studied exclusive nonleptonic and semileptonic decays of the $B_c$ meson within a relativistic constituent quark model developed by them. In their recent work [8], they have calculated the nonleptonic decays with one of the final state being pure $c\bar{c}$. They predict $B(B_c^- \to \pi^- \chi_{c2})$ and $B(B_c^- \to \rho^- \chi_{c2})$ as $4.6\times10^{-4}$ and $1.2\times10^{-3}$ respectively, which are large as compare to present results. Similarly, in another recent work [5] the same decays have been quoted with the branching ratios ($B(B_c^- \to \pi^- \chi_{c2}) = 2.2\times10^{-4}$ and $B(B_c^- \to \rho^- \chi_{c2}) = 6.5\times10^{-4}$), that are of the same order of magnitude as the compared to present work. It has also been observed that the largest numerical values of branching ratios $B_c \to PT/VT$ are of the same order as those of some $B_c \to PP/PV/VV$ decay modes [4-12]. In $B$ meson decays, the experimental data favors constructive interference, in contrast to the charm meson sector, between the color favored and color suppressed diagrams, thereby yielding $a_1 = 1.10 \pm 0.08$ and $a_2 = 0.20 \pm 0.02$. Our results remain unaffected from interference of $a_1$ (color favored) and $a_2$ (color suppressed).

## V. SUMMARY AND CONCLUSIONS

In this paper, the two-body hadronic weak decays of $B_c$ meson to pseudoscalar/vector and tensor mesons in charm changing as well as bottom changing decay modes, employing the ISGW II model, are studied. However, the charm changing decay modes ($b$ spectator) are kinematically forbidden in case of the decays involving one vector meson in the final state. Various form factors for $B_c \to T$ transition are obtained using the Isgur-Scora-Grienstein-Wise (ISGW II) quark model. Consequently, branching ratios of $B_c \to PT$ and $B_c \to VT$ for various Cabibbo-favored decay modes are calculated. Conclusions are as follows.

i) Dominant decay modes for $B_c \to PT$ are $B(B_c^+ \to \pi^+ B_{s2}^0)$ = 3.0×10$^{-4}$, $B(B_c^- \to D_s^- \chi_{c2})$ = 3.2×10$^{-4}$ and $B(B_c^- \to \pi^- \chi_{c2})$ = 2.0×10$^{-4}$. For $B_c \to VT$ decays, $B(B_c^- \to D_s^{*-} \chi_{c2})$ = 8.8×10$^{-4}$ and $B(B_c^- \to \rho^- \chi_{c2})$ = 2.4×10$^{-4}$, are dominant decays. Branching ratios for these decays seem to be within the reach of current experiments. Observation of these processes in the $B_c$ experiments such as Belle, Babar, BTeV, LHC and so on will be crucial in testing the ISGW quark model as well as validity of the factorization scheme.

ii) In contrast to the charm meson sector, the experimental data of $B$ meson decays favor the constructive interference between color favored and color suppressed diagrams, giving $a_1 = 1.10 \pm 0.08$ and $a_2 = 0.20 \pm 0.02$. In the present analysis, the decay amplitude is proportional to only one QCD coefficient either $a_1$(for color favored diagram) or $a_2$(for color suppressed diagram), therefore these results remain unaffected from interference of $a_1$ and $a_2$.

## ACKNOWLEDGEMENT

Author is thankful to Prof. R.C. Verma and Rohit Dhir for discussions, suggestions and careful reading of the manuscript and also acknowledges the financial assistance from the University Grant Commission, New Delhi.

**Table I. The values of parameter $\beta$ for $s$-wave and $p$-wave mesons in the ISGW II quark model**

| Quark content | $u\bar{d}$ | $u\bar{s}$ | $s\bar{s}$ | $c\bar{u}$ | $c\bar{s}$ | $u\bar{b}$ | $s\bar{b}$ | $c\bar{c}$ | $b\bar{c}$ |
|---|---|---|---|---|---|---|---|---|---|
| $\beta_s$ (GeV) | 0.41 | 0.44 | 0.53 | 0.45 | 0.56 | 0.43 | 0.54 | 0.88 | 0.92 |
| $\beta_p$ (GeV) | 0.28 | 0.30 | 0.33 | 0.33 | 0.38 | 0.35 | 0.41 | 0.52 | 0.60 |

**Table II. Form factors of $B_c \to T$ transition at $q^2 = t_m$ in the ISGW II quark model**

| Modes | Transition | $h$ | $k$ | $b_+$ | $b_-$ |
|---|---|---|---|---|---|
| $\Delta b = 1, \Delta C = 0, \Delta S = -1$ | $B_c \to D_2$ | 0.017 | 0.556 | -0.008 | 0.011 |
| | $B_c \to D_{s2}$ | 0.019 | 0.739 | -0.011 | 0.014 |
| $\Delta b = 0, \Delta C = -1, \Delta S = -1$ | $B_c \to B_2$ | 0.100 | 2.722 | -0.034 | 0.148 |
| | $B_c \to B_{s2}$ | 0.119 | 3.632 | -0.049 | 0.165 |
| $\Delta b = 1, \Delta C = 1, \Delta S = 0$ | $B_c \to \chi_{c2}$ | 0.023 | 1.411 | -0.017 | 0.019 |

**Table III. Decay amplitudes of $B_c \to PT$ decays for Cabibbo-favored modes in Charm changing and Bottom changing decays**

| Decays | Amplitude |
|---|---|
| $\Delta b = 0, \Delta C = -1, \Delta S = -1$ | |
| $B_c^+ \to \pi^+ B_{s2}^0$ | $a_1 f_\pi F^{B_c \to B_{s2}}(m_\pi^2) V_{cs} V_{ud}^*$ |
| $B_c^+ \to \overline{K}^0 B_2^+$ | $a_2 f_K F^{B_c \to B_2}(m_K^2) V_{cs} V_{ud}^*$ |
| $\Delta b = 1, \Delta C = 1, \Delta S = 0$ | |
| $B_c^- \to \pi^- \chi_{c2}$ | $a_1 f_\pi F^{B_c \to \chi_{c2}}(m_\pi^2) V_{cb} V_{ud}^*$ |
| $B_c^- \to D^0 D_2^-$ | $a_2 f_D F^{B_c \to D_2}(m_D^2) V_{cb} V_{ud}^*$ |
| $\Delta b = 1, \Delta C = 0, \Delta S = -1$ | |
| $B_c^- \to \pi^0 D_{s2}^-$ | $\frac{1}{\sqrt{2}} a_2 f_\pi F^{B_c \to D_{s2}}(m_\pi^2) V_{ub} V_{us}^*$ |
| $B_c^- \to \eta D_{s2}^-$ | $\frac{1}{\sqrt{2}} a_2 f_\eta \sin\phi_P F^{B_c \to D_{s2}}(m_\eta^2) V_{ub} V_{us}^*$ |
| $B_c^- \to K^- \overline{D}_2^0$ | $a_1 f_K F^{B_c \to D_2}(m_K^2) V_{ub} V_{us}^*$ |
| $B_c^- \to \eta' D_{s2}^-$ | $\frac{1}{\sqrt{2}} a_2 f_{\eta'} \cos\phi_P F^{B_c \to D_{s2}}(m_{\eta'}^2) V_{ub} V_{us}^*$ |
| $B_c^- \to D_s^- \chi_{c2}$ | $a_1 f_{D_s} F^{B_c \to \chi_{c2}}(m_{D_s}^2) V_{cb} V_{cs}^*$ |
| $B_c^- \to \eta_c D_{s2}^-$ | $a_2 f_{\eta_c} F^{B_c \to D_{s2}}(m_{\eta_c}^2) V_{cb} V_{cs}^*$ |

**Table IV. Decay amplitudes of $B_c \to VT$ decays for Cabibbo-favored modes in Bottom changing decay**

| Decays | Amplitude |
|---|---|
| $\Delta b = 1, \Delta C = 1, \Delta S = 0$ | |
| $B_c^- \to \rho^- \chi_{c2}$ | $a_1 m_\rho f_\rho F_{\alpha\beta}^{B_c \to \chi_{c2}} V_{cb} V_{ud}^*$ |
| $B_c^- \to D^{*0} D_2^-$ | $a_2 m_{D^*} f_{D^*} F_{\alpha\beta}^{B_c \to D_2} V_{cb} V_{ud}^*$ |
| $\Delta b = 1, \Delta C = 0, \Delta S = -1$ | |
| $B_c^- \to \rho^0 D_{s2}^-$ | $\frac{1}{\sqrt{2}} a_2 m_\rho f_\rho F_{\alpha\beta}^{B_c \to D_{s2}} V_{ub} V_{us}^*$ |
| $B_c^- \to \omega D_{s2}^-$ | $\frac{1}{\sqrt{2}} a_2 m_\omega f_\omega \sin\phi_V F_{\alpha\beta}^{B_c \to D_{s2}} V_{ub} V_{us}^*$ |
| $B_c^- \to K^{*-} \bar{D}_2^0$ | $a_1 m_{K^*} f_{K^*} F_{\alpha\beta}^{B_c \to D_2} V_{ub} V_{us}^*$ |
| $B_c^- \to \phi D_{s2}^-$ | $\frac{1}{\sqrt{2}} a_2 m_\phi f_\phi \cos\phi_V F_{\alpha\beta}^{B_c \to D_{s2}} V_{ub} V_{us}^*$ |
| $B_c^- \to D_s^{*-} \chi_{c2}$ | $a_1 m_{D_s^*} f_{D_s^*} F_{\alpha\beta}^{B_c \to \chi_{c2}} V_{cb} V_{cs}^*$ |
| $B_c^- \to J/\psi D_{s2}^-$ | $a_2 m_{j/\psi} f_{j/\psi} F_{\alpha\beta}^{B_c \to D_{s2}} V_{cb} V_{cs}^*$ |

**Table V. Branching ratios (in %) of semileptonic $B_c$ decays**

| Decays | Branching ratios (%) | | | | | | |
|---|---|---|---|---|---|---|---|
| | ISGW II | [5] | [6] | [7] | [8] | [9] | [10] |
| $B_c^+ \to \eta_c e \bar{\nu}_e$ | 0.66 | 0.48 | 0.15 | 0.40 | 0.81 | 0.59 | 0.51 |
| $B_c^+ \to J/\psi e \bar{\nu}_e$ | 1.07 | 1.54 | 1.47 | 1.21 | 2.07 | 1.20 | 1.44 |
| $B_c^+ \to B_s e \bar{\nu}_e$ | 0.96 | 1.06 | 0.8 | 0.82 | 1.10 | 0.99 | 0.92 |
| $B_c^+ \to B_s^* e \bar{\nu}_e$ | 1.17 | 2.35 | 2.3 | 1.71 | 2.37 | 2.30 | 1.41 |

**Table VI. Branching ratios of $B_c \to PT$ decays for Cabibbo-favored modes in Charm changing and Bottom changing decays**

| Decays | Branching ratios | | |
|---|---|---|---|
| | This work | [13] | [14] |
| $\Delta b = 0, \Delta C = -1, \Delta S = -1$ | | | |
| $B_c^+ \to \pi^+ B_{s2}^0$ | $3.0\times10^{-4}$ | - | $2.01\times10^{-4}$ |
| $B_c^+ \to \overline{K}^0 B_2^+$ | $1.0\times10^{-5}$ | - | $4.22\times10^{-6}$ |
| $\Delta b = 1, \Delta C = 1, \Delta S = 0$ | | | |
| $B_c^- \to \pi^- \chi_{c2}$ | $2.0\times10^{-4}$ | $2.48\times10^{-4}$ | $7.5\times10^{-5}$ |
| $B_c^- \to D^0 D_2^-$ | $4.0\times10^{-6}$ | - | $6.26\times10^{-8}$ |
| $\Delta b = 1, \Delta C = 0, \Delta S = -1$ | | | |
| $B_c^- \to \pi^0 D_{s2}^-$ | $6.8\times10^{-10}$ | - | $1.99\times10^{-11}$ |
| $B_c^- \to \eta D_{s2}^-$ | $3.6\times10^{-10}$ | - | $2.51\times10^{-12}$ |
| $B_c^- \to K^- \overline{D}_2^0$ | $1.6\times10^{-8}$ | - | $1.43\times10^{-10}$ |
| $B_c^- \to \eta' D_{s2}^-$ | $3.1\times10^{-10}$ | - | $1.74\times10^{-11}$ |
| $B_c^- \to D_s^- \chi_{c2}$ | $3.2\times10^{-4}$ | $4.54\times10^{-4}$ | $1.54\times10^{-4}$ |
| $B_c^- \to \eta_c D_{s2}^-$ | $1.4\times10^{-5}$ | - | $1.4\times10^{-5}$ |

**Table VII. Branching ratios of $B_c \to VT$ decays for Cabibbo-favored modes in Bottom changing decays**

| Decays | Branching ratios | | |
|---|---|---|---|
| | **This work** | **[13]** | **[14]** |
| $\Delta b$ =1, $\Delta C$ = 1, $\Delta S$ = 0 | | | |
| $B_c^- \to \rho^- \chi_{c2}$ | $2.4\times10^{-4}$ | $5.18\times10^{-4}$ | $2.38\times10^{-4}$ |
| $B_c^- \to D^{*0} D_2^-$ | $8.9\times10^{-6}$ | - | $3.42\times10^{-7}$ |
| $\Delta b$ =1, $\Delta$C = 0, $\Delta S$ = -1 | | | |
| $B_c^- \to \rho^0 D_{s2}^-$ | $8.4\times10^{-10}$ | - | $7.91\times10^{-11}$ |
| $B_c^- \to \omega D_{s2}^-$ | $5.3\times10^{-11}$ | - | $3.94\times10^{-11}$ |
| $B_c^- \to K^{*-} \bar{D}_2^0$ | $1.6\times10^{-8}$ | - | $4.52\times10^{-10}$ |
| $B_c^- \to \phi D_{s2}^-$ | $1.2\times10^{-9}$ | - | $4.81\times10^{-11}$ |
| $B_c^- \to D_s^{*-} \chi_{c2}$ | $8.8\times10^{-4}$ | $2.4\times10^{-3}$ | $5.25\times10^{-4}$ |
| $B_c^- \to J/\psi D_{s2}^-$ | $9.6\times10^{-5}$ | - | $2.06\times10^{-5}$ |